К. Ефимов[1]

# ИДЕНТИФИКАЦИЯ КАРТЕЛЕЙ НА ЭЛЕКТРОННЫХ АУКЦИОНАХ ГОСЗАКУПОК[2]

Исследование, направленное на выявление картельных сговоров, включало анализ решений Федеральной антимонопольной службы России и данных об электронных аукционах. В результате была разработана модель машинного обучения, которая с точностью 91% предсказывает признаки сговора между участниками торгов на основе их истории после деления 40 аукционов на тестовую и обучающую выборки в соотношении 30/70. Декомпозиция модели с использованием вектора Шепли позволила интерпретировать процесс принятия решений. Также было изучено поведение "честных" компаний на аукционах, подтвержденное независимой симуляционной проверкой.

Ключевые слова: картели, электронные аукционы, госзакупки.

В 2023 году через систему госзакупок было заключено порядка 2.4 миллиона контрактов, общая стоимость которых достигла довольно существенной суммы в 10.6 трлн рублей. Этот огромный рынок обладает значительным потенциалом для антиконкурентных соглашений. Некоторые компании могут вступать в недобросовестные картельные сговоры, которые ограничивают конкуренцию на рынке с целью увеличения своей прибыли. Такие соглашения приводят к повышению цен на заключение контрактов, в конечном итоге увеличивают стоимость товаров и услуг для граждан, что недопустимо.

Картельные соглашения запрещены в большинстве стран мира антимонопольных законодательством, как соглашения, негативно отражающиеся на рынке. Несмотря на это, картели очень часто встречаются при проведении госзакупок во многих странах, включая Россию. Федеральная антимонопольная служба (ФАС) активно занимается поиском таких сговоров, развернув обширную правоприменительную практику. Однако, огромный объем

[1] Ефимов Константин Дмитриевич- аспирант, Национальный исследовательский университет Высшая школа экономики (Санкт- Петербург, 190121)

системы госзакупок создаёт сложности для эффективного контроля соблюдения антимонопольных законов.

Процесс автоматизации и систематизации выявления картельных сговоров мог бы существенно снизить убытки, понесенные государством от такой деятельности. Машинное обучение может помочь с выполнением этой задачи. Оно способно автоматизировать процесс поиска картелей, облегчая тем самым задачу контроля.

Главный исследовательский вопрос, поднимаемый в этой работе- можно ли обнаружить сговор с помощью методов машинного обучения? Гипотеза- модели машинного обучения могут обнаружить сговоры на аукционах лучше случайного угадывания.

С развитием цифровизации системы госзакупок, внедрение такой модели становится всё более достижимым. Настоящая работа затрагивает обсуждение текущих методов поиска картельных сговоров, представленных в научной литературе, предлагает новую методологию поиска, предложенную автором, оценивает ее потенциальную эффективность и изучает уроки, которые можно извлечь, анализируя процесс принятия решений в данной области.

## АНАЛИЗ АЛГОРИТМОВ ПОИСКА КАРТЕЛЕЙ

Методы машинного обучения позволяют анализировать большие объемы данных и выявлять сложные шаблоны поведения участников. Различные исследователи

Исследователи пробовали применять регрессионный анализ для поиска картелей среди аукционов на укладку асфальта в разных префектурах страны [1]. С помощью анализа удалось установить, что при появлении на аукционе участника из другой префектуры цена на аукционе резко снижалась, что объясняется объявлением ценовой войны картелем иногородней компании. Это необычный результат, показывающий, что снижение цены не всегда говорит об отсутствии картеля на рынке.

В следующей работе авторы собрали данные о швейцарских аукционах госзакупок, где участвовали картели [2]. Построили для каждой пары фирм с аукционов графики поставленных ими ставок на всех аукционах, в которых они присутствовали. Получившиеся графики разметили, где находились ставки картелей и обучили на графиках CNN- нейронную сеть для распознавания картинок, которая смогла с точностью более 90% правильно предсказать наличие картеля среди двух фирм по графику общих ставок на аукционах. Это уникальный подход использования нейросети для распознавания картинок в такой задаче.

Использовались и самые современный подход в машинном обучении для поиска картелей: частичное обучение с учителем [3]. Это задача в машинном обучении, смежная с задачей классификации и кластеризации. Ученые исследуют аукционы на предмет неконкурентности, для этого используют примеры аукционов с несколькими участниками как конкурентный класс, а аукционы с одним участником как неразмеченный класс, ведь среди них могут быть как случаи проявления картелей, так и ситуации, когда единственный участник ведется себя как в конкурентной среде. Ведь по заявлениям авторов фирмы не знают количество соперников на аукционе. Алгоритм PU- learning ( positive- unlabeled) позволил авторам выделить группу аукционов с предполагаемым наличием монополизации и картельного виляния.

Существует и похожая статья с использованием PU- learning стратегии [4]. Авторы рассуждают, что в некоторых случаях может образоваться картель между организатором закупки и одним из его участников. Организатор может сливать информацию о ставках конкурентов своему фавориту. Такое незаконное поведение нужно искать среди победителей аукционов, кто ставил ставку последним. Но об этом поведении могут знать и сами участники аукционов. Честные участники, желающие предотвратить раскрытие информации о своей ставке, могут также начать ставить ставку в последние минуты аукциона. Используя PU- learning, исследователи установили долю картельных аукционов на уровне 9% от выборки в 600000 аукционов.

Большое развитие получили и работы о комбинации нескольких удачных подходов поиска картелей с машинным обучением. В одной из таких работ авторы указывают на возможность комбинации метода красных флагов и машинного обучения [5]. Суть метода красных флагов заключается в возможности пометить подозрительную активность фирм на аукционе посредством различных статистических показателей или информации о документации аукциона. Так, чем больше красных флагов у конкретного аукциона, тем подозрительнее на наличие нарушений в нем. А что если применить этот метод с машинным обучением? Авторы выбрали 18 красных флагов на аукционах, связанных с полнотой и прозрачностью организации закупки и попробовали построить модель машинного обучения для предсказания наличия картеля на итальянских тендерах. Переменные включали в себя такие показатели как наличие публичного объявления о тендере, наличие разъяснений о проведении тендера, время на подачу заявки, регламентацию субподрядов и другие. Построенная модель машинного обучения не обладает высокой предсказательной способностью, но показала явный потенциал некоторых индикаторов в качестве реальных красных флагов.

Исследователи изучали и оценку эффективности использования методов машинного обучения для выявления картелей на примере швейцарский аукционах для госзакупок [6]. Авторы смогли достичь 84% точности обнаружения картелей.

Методы машинного обучения могут быть применены не только на аукционах, но и на обычных рынках [7].Так, исследователи проводили анализ рынка бензина на автозаправок в одной из стран. Авторы показывают, что в моменты доминирования на рынке картельного соглашения, дисперсия цены во времени была значительно меньше, чем у рынка бензина в свободное от картеля времени, что было доказано с помощью включения этого статистического показателя в модель машинного обучения.

Использование похожих статистических показателей на аукционах позволяет получить больше ценной информации об аукционах. В данной работе берут некоторое количество ставок из каждого аукциона, анализируя такие показатели как среднее, коэффициент вариации, дисперсия ставок с аукционов [8]. После построения модели исследователи получили робастные оценки значимости каждого из показателей в лассо регрессии.

Существует большое количество методов оценки качества работы моделей машинного обучения. В представленной работе авторы обсуждают, что важнее, избежать ложноположительных или ложноотрицательных результатов [9]. Они обращают внимание на возможность небинарной классификации аукционов: если вероятность картеля предсказанная алгоритмом составляет от 0.5 до 0.7, то он может помещаться в класс условно подозрительных, если же более, то отмечается как крайне подозрительный. Крайне подозрительные картели отмечаются как картельные примеры, в то время как просто подозрительные на картельные примеры отправляются на дополнительную проверку следующей модели, решающей уже задачу бинарной классификации. Такая стратегия уменьшает количество ложно положительных результатов. Модель использовала лишь ставки участников в качестве признаков и достигла результатов в 80%. Похожие наработки авторы публиковали и в [10]

Исследователи придумали совместить подходы обучения с учителем и обучения без учителя в машинном обучении [11]. Сначала они кластеризуют данные, после чего стараются предсказать полученные кластеры с помощью задачи классификации, зная какие классы представляют из себя скопления картельных аукционов. Точность такого алгоритма составила 80%.

Наличие механизмов поиска картелей и санкции их участникам привели к изменению стратегии организации и координации картелей. Авторы работы показывают, что в случае, если существуют соответствующие институты контроля, фирмы стараются использовать инсинуации и непрямые намеки между собой для создания сговора на рынках, что усложняет их поиск. [12]

На примере бразильских госзакупок исследователи обсуждают процесс создания системы из нескольких автоматизированных агентов для сбора информации о проведенных тендерах [13]. Авторы показывают возможную функциональную структуру и принцип работы такого алгоритма, способ хранения данных по результатам работы такого алгоритма и сравнивают его с классическими методами дата майнинга.

Отечественные исследователи применяли опыт использования машинного обучения в Российской системе госзакупок [14]. Авторы этой работы применяют статистические показатели в различных моделях машинного обучения для поиска картелей в отечественных госзакупках. Они анализируют такие показатели как начальная цена контракта, минимальная разница во времени между подаче ставок участниками и другие непрерывные показатели для построения модели машинного обучения, что позволяет находить картели с точностью 87%

Вывод из обзора литературы: применение методов машинного обучения активно развито в литературе, но некоторые подходы требуют большого объема данных, в то время как другие подходят не для всех аукционов. Существует различие между закрытым аукционом первой цены и открытым английским аукционом. Второй тип предполагает динамику внутри самого аукциона в виде возможности видеть текущую цену на аукционе и подавать несколько ставок от имени одной организации. Большинство из рассмотренных статей фокусируются только на закрытом аукционе или представляют открытые аукционы последними ставками каждого из игроков, теряя часть информации.

# ДИЗАЙН ИССЛЕДОВАНИЯ

После рассмотрения правоприменительной практики обратимся к дизайну исследования. Данная работа фокусируется на открытых аукционах в государственных закупках. Стратегия обнаружения сговора заключается в предложении в наличии фундаментальных различий поведения независимых фирм и картелей, вытекающей из- за различия задач максимизации прибыли. Стратегия обнаружения сговора предполагает установление ожидаемого поведения честной фирмы теоретическим способом, после чего можно было бы использовать сравнивание поведения фирм с ожидаемым стандартом.

Для начала, опишем обычные правила на самом аукционе. Каждая из фирм может подать ставку на аукционе, предлагая свое снижение цены контракта, причем снижать цену можно лишь в допустимых пределах, от 0.5% до 5% от начальной цены. Так, на лот стоимостью 100 руб за единицу товара можно подавать предложения от 99.5 руб до 95 руб за единицу товара. Каждая фирма видит текущее снижение цены и имеет возможность предложить следующее снижение цены от 0.5% до 5%. Если в течении 10 минут ставка не подается, то последняя отправленная ставка считается победившей.

Как честной фирме в таких условиях выбрать оптимальное снижение цены своей ставкой? Можно доказать, что лучшей идеей будет снижать ставку каждый раз на минимально возможную величину, так как если предлагать большую ставку, можно случайно перепрыгнуть точку бегства конкурента. Вернемся к примеру аукциона, на котором разыгрывается контракт на поставку какого то товара стоимостью 100 руб за штуку ( обычно в контракте указывается общая сумма, а не за единицу продукции, но это сейчас не играет роли). Пусть у фирмы 1 точка бегства на уровне 60 руб, а у 2 фирмы 70 руб. Пусть текущая цена после ставки фирмы 2 стала 70 рублей 1 копейка. Если предложить любое снижение цены от лица первой фирмы, то она выиграет, но вот выручка от контракта будет максимальной только в случае предложения наименьшего снижения цены, так как тогда текущая цена будет как можно ближе к точке бегства фирмы. Этот вывод очень похож на результаты классической модели олигополии Бертрана, в которой фирма с меньшими издержками предлагает цену равную издержкам второй фирмы минус бесконечно малая величина.

Чтобы доказать приведенные рассуждения, можно провести симуляционный процесс тендерного аукциона. Симуляция является сложной структурой, состоящей из нескольких объектов: популяции агентов и аукциона. Популяция агентов является проекцией возможного пространства стратегий, которые реальные фирмы могут представлять в жизни, причем используемые стратегии могут меняться с течением времени. Из популяции агентов случайным образом выбираются два игрока для участия в аукционе. У каждого из игроков есть свои издержки, распределенные равномерно от 0 до 1, Н аукционе, стартовая цена которого нормирована и всегда равна 1, игроки по очереди делают ставки. Фирма, которая будет ходить первой, выбирается случайным образом.Фирма делает ставку, если после сделанной ставки текущая цена на аукционе не меньше чем издержки фирмы. В противном случае фирма не дает ставку и считается проигравшей. Для победителя считается величина прибыли, это текущая цена минус издержки фирмы. Эта величина всегда принадлежит отрезку от 0 до 1. Проигравший игрок не может воссоздать технические условия, чтобы получить уменьшение издержек, например, до уровня победителя, но может наблюдать его стратегию подачи ставок на аукционах. Прибыль участника представляет собой некоторую привлекательность стратегии победителя и равна вероятности конвертации стратегии агента проигравшего в стратегию победителя.

## СИМУЛЯЦИОННЫЙ ПРОЦЕСС.

Результаты симуляции представлены на графиках 1,2 и 3. Всего в популяции есть три типа игроков: агрессивные игроки, всегда снижающие цену на 5%, пассивные игроки, снижающие цену на 0.5% и случайные игроки, снижающие цену случайным образом от 0.5% до 5%. Было проведено 3 симуляции, в каждой участники сыграли 20000 аукционов ( кроме последней, в ней было 50000) Во всех симуляциях размер популяции был равен 100. Различались лишь стартовые точки. В первой симуляции было равное количество каждого из агентов, во второй симуляции пассивных агентов было всего 10% в третьей- 3%.

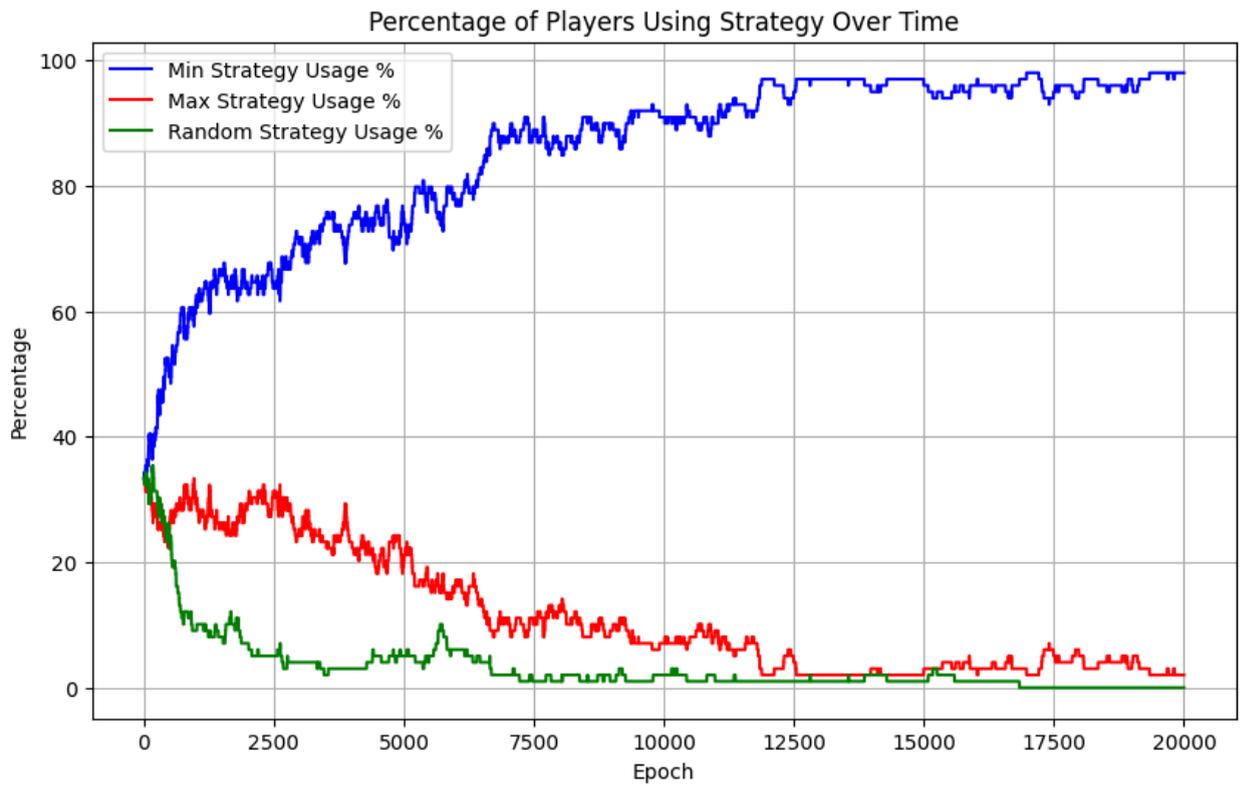

График 1. Доли различных типов агентов в популяции с течением времени. Источник: расчеты автора

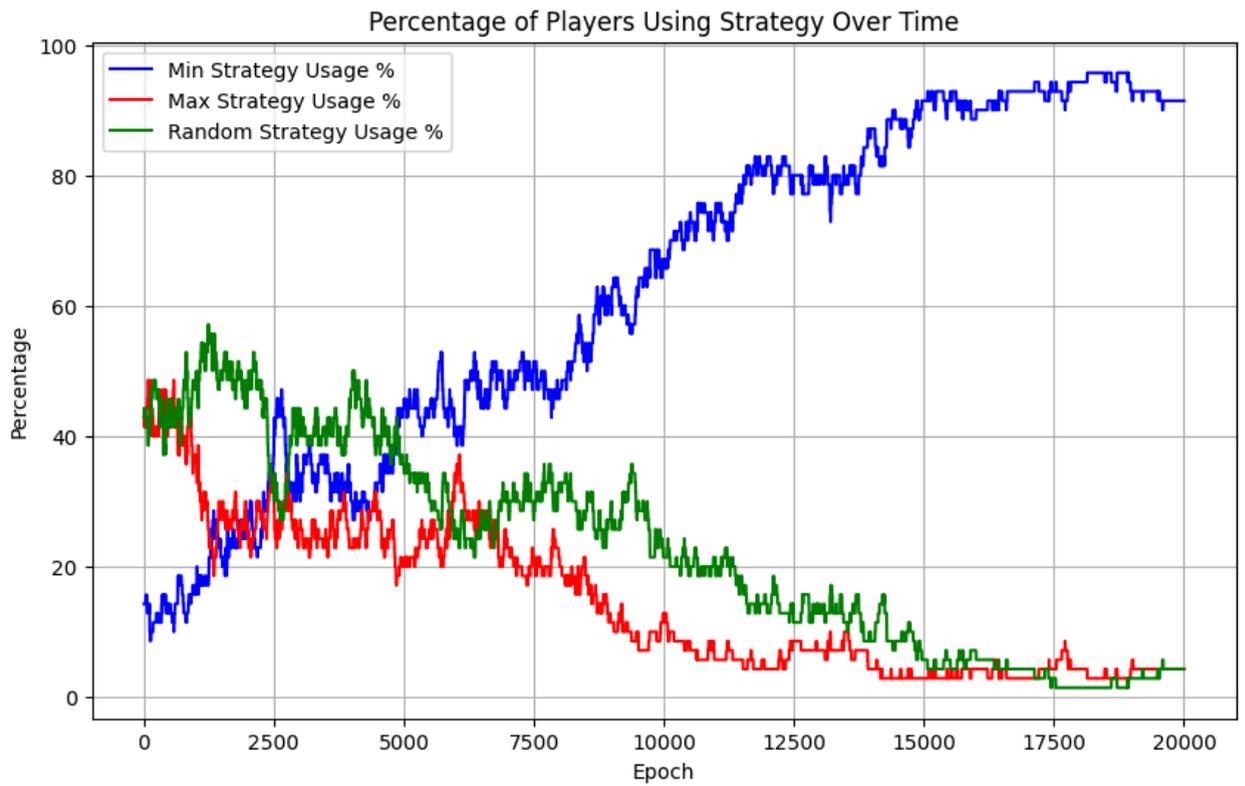

График 2. Доли различных типов агентов в популяции с течением времени. Источник: расчеты автора

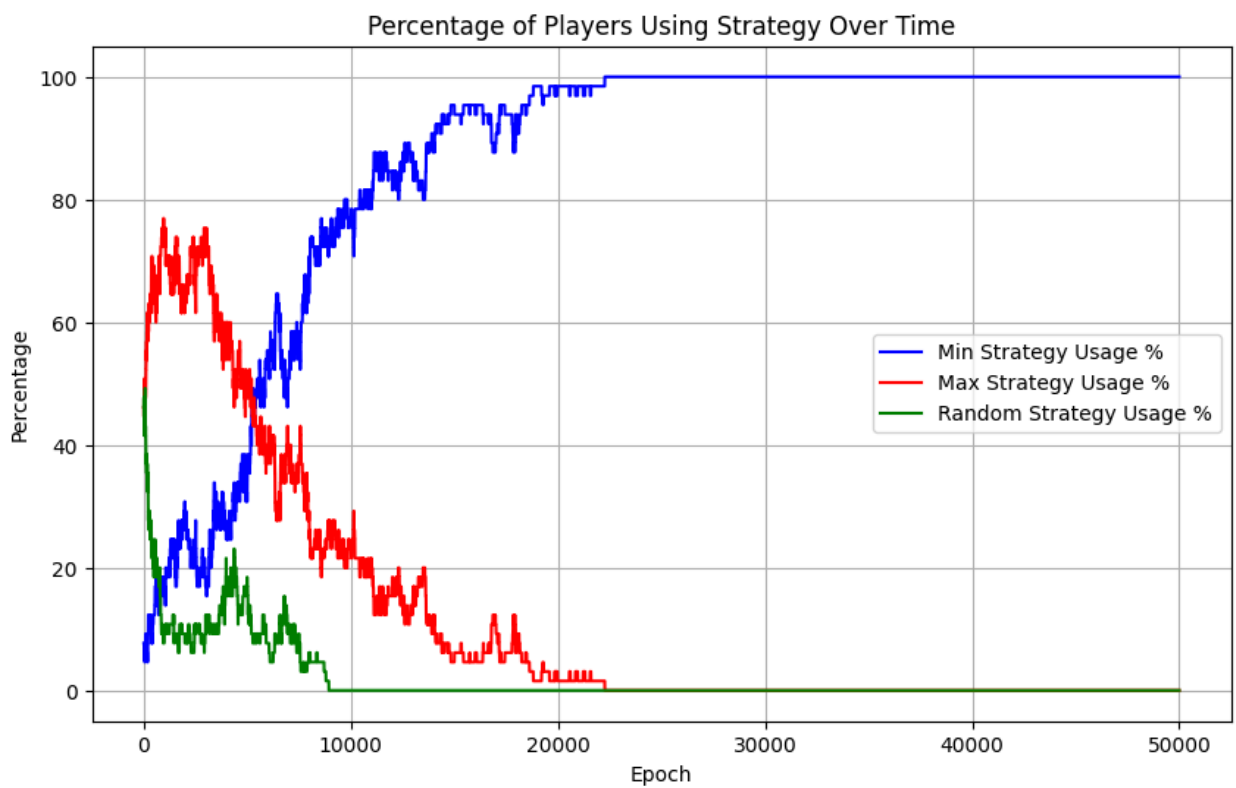

График 3. Доли различных типов агентов в популяции с течением времени. Источник: расчеты автора.

Можно заметить, что в каждой из симуляции количество пассивных игроков увеличивалось, конвергенция достигалась при полном доминировании агентов, снижающих цену на наименьшее возможное значение. Из этого следует, что в такой популяции останутся только те агенты, что снижают цену на наименьшее возможное значение.

# МАШИННОЕ ОБУЧЕНИЕ

Изначальные данные включают в себя данные об истории торгов с 89000 аукционов, проведенных в период с 2016 по 2020 года по 223- ФЗ РФ. Для определения картельных аукционов с федерального сайта ФАС России были собраны тексты более 1100 дел о создании картелей. В текстах указаны номера аукционов, в которых были найдены картели. После сопоставления 89000 собранных аукционов и текстов дел, было обнаружено 20 аукционов, в которых было доказано наличие картелей. В виду анализа истории ставок на аукционе, данные представляют из себя 20 временных рядов разной длины. В общей выборки количество ставок на аукционе разнилось от 1 до 234. Чтобы привести ряды к одной длине, все ряды были обрезаны до 11 первых ставок, а все ряды с меньшим размером дополнены последней ставкой до 11. Автор работы пробовал и другие длины рядов, существенной разницы в качестве модели обнаружено не было. В качестве примеров честных аукционов были взяты 20 случайных аукционов для сбалансированности выборки. Таким образом, итоговая выборка аукционов состоит из 40 историй торгов, по 11 ставок в каждом. В качестве модели была взята модель градиентного бустинга. Стратегия оценивания предполагает выдержанный подход для определения метрики качества модели. Существует явная проблема малого размера выборки, которая может привести к искажению оценки качества модели из- за искажения модели. Существуют исследования, описывающие работу с оценкой качества модели на маленьких выборках. В частности, в работе, посвященной малым выборкам в машинном обучении, советуют использовать кросс-валидацию внутри деления на тренировочную и тестовую выборку [15]. Это приведет к получению робастных оценок качества модели. Разделим выборку на тренировочную и тестовую в соотношении 70 на 30, внутри тренировочной выборки проведем 5 фолдовую кросс- валидацию с подбором гиперпараметров. Результаты модели с лучшей точностью представлены в таблице 1.

|  | Предсказанный класс | |
|---|---|---|
| Истинный класс | Честный | Картель |
| Честный | 41% | 0 |
| Картель | 9% | 50% |

Таблица 1. Результаты градиентного бустинга. Источник: расчеты автора.

91% аукционов из тестовой выборки были верно определены. С учетом того, что модель опирается лишь на некоторое количество первых ставок из истории торгов, она не является ресурсозатратно и может быть применена сразу после окончания аукциона.

# ВЕКТОР ШЕПЛИ

С развитием новых технологий можно получить не только модель машинного обучения, но можно получить не только саму модель, но и полезные выводы о закономерностях ее работы. Одни из самых продвинутых исследований, в частности, пытаются понять процесс принятия решений нейронными сетями, какие именно паттерны они находят в данных ,что позволяет им так хорошо предсказывать классы.

В нашем случае нам поможет библиотека shap, реализующая концепцию вектора шепли в питоне. Вектор Шепли является теоретико игровой концепцией, представляющей вектор справедливого распределения общего выигрыша некоторой большой коалиции игроков. Выплата каждого игрока зависит от результата коалиции с ним и результатов коалиции без этого игрока. Таким образом под справедливостью представляется вклад игрока в общий результат команды. Если под игроками представлять каждую из характеристик набора данных, а результатом коалиции считать точность предсказания, то получим интерпретацию вектора шепли для модели машинного обучения.

Посмотрим несколько конкретных примеров предсказанных аукционов из тестовой выборки. Как именно модель модель принимала решения на основе представленных данных? График аукциона 1 из тестовой выборки представлен на графике 4

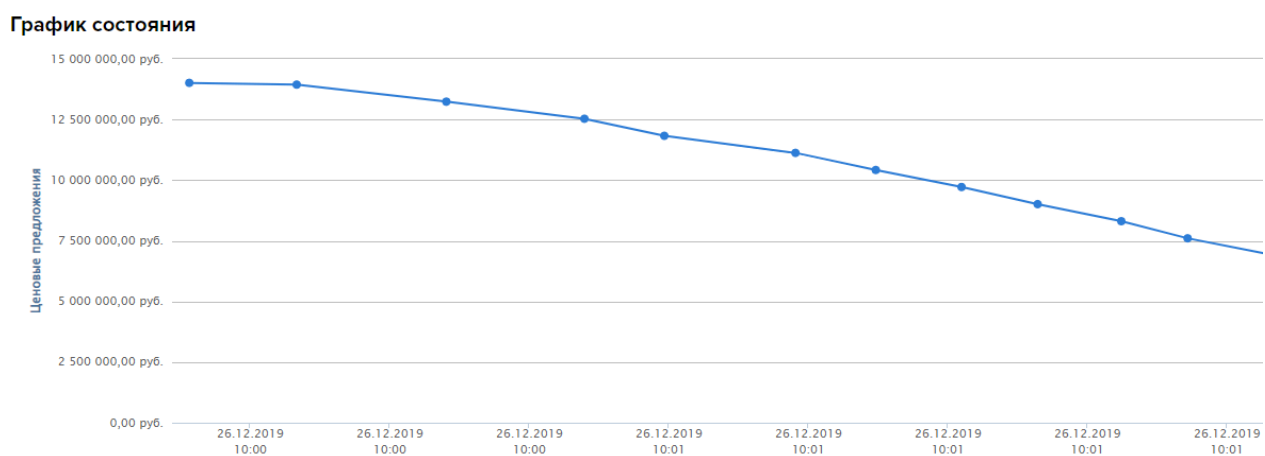

График 4. Ход торгов на аукционе с реестровым номером № 31908675706. Источник: РТС-Тендер.

Является ли данных аукцион конкурентным? Цена за 1 минуту упала с около 15 млн до 7.5 млн рублей, что превышает отметку в 50% снижение. Можно с уверенностью сказать, что такой аукцион проводился в конкурентной среде. Но это будет заблуждением. Обратимся к материалам дела №077/01/11-15531/2021 Московского УФАС России: "В ходе аукциона с реестровым № 31908675706, проведенном 26.12.2019, поочередно подавались ценовые предложения ООО «Платинум Плюс» (-1 %), ООО «Строймонтаж» (-81%), ООО «Вега» (-76,00%), ООО «Компания «Платинум» (-0,5%). При этом ООО «Вега» и ООО «Строймонтаж» подавали ценовые предложения с разницей во времени в несколько секунд. Указанные действия привели к быстрому снижению НМЦК до 2 675 336,24 руб., то есть до -81% от НМЦК". По результатам рассмотрения дела данные компании были признаны заключившими устное картельное соглашение. Поведение фирм описывается следующим образом: "Из вышеизложенного следует, что ООО «Компания Платинум», ООО «Вега», ООО «Строймонтаж», ООО «Платинум Плюс» была реализована антиконкурентная модель поведения, именуемая «таран», при которой добросовестные участники торгов утрачивают экономический интерес в торгах, вследствие значительного снижения начальной максимальной цены контракта сговорившимися участниками, у которых нет реального намерения заключить контракт, а есть только цель введения в заблуждение относительно размера снижения цены." Как градиентный бустинг оценил данный аукцион? Обратимся к графику 5, показывающему разложение вектора Шепли.

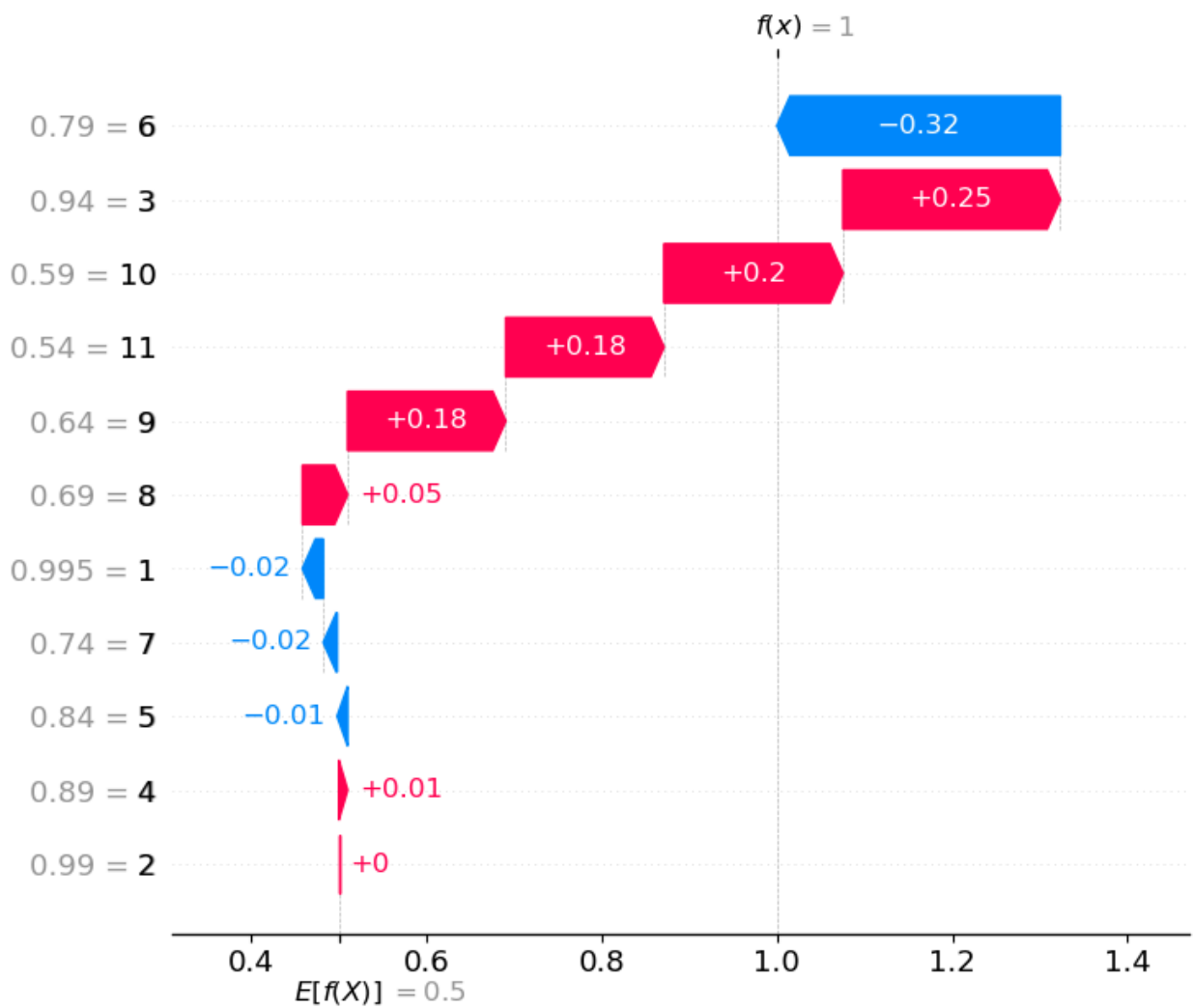

График 5. Интерпретация процесса решения градиентного бустинга. Ход торгов на аукционе с реестровым номером № 31908675706. Источник: РТС- Тендер.

На оси X представлено математическое ожидание аукциона оказаться картелем, стартующее в точке 0.5. По оси y отмечены порядковые номера ставок на данном аукционе, от самой незначимой до самой важной по мнению градиентного бустинга. На самом графике показаны смещения матожидания вероятности аукциона оказаться картелем и их величина. Как можно заметить, первая ставка, поданная на аукционе была равна 0.995, то есть участник предложил наименьшее возможное снижение цены одной ставкой. Тоже самое верно и для второй ставки. А вот третья ставка была равна 94%, снижение цены относительно второй ставки (95%) составило 5%. Увидев это, модель увеличила вероятность появления картеля сразу на 25%. Максимальное снижение цены показалось модели довольно подозрительным, и, как мы узнали из материалов дела, далеко не зря. Похожий пример представлен на графике 6.

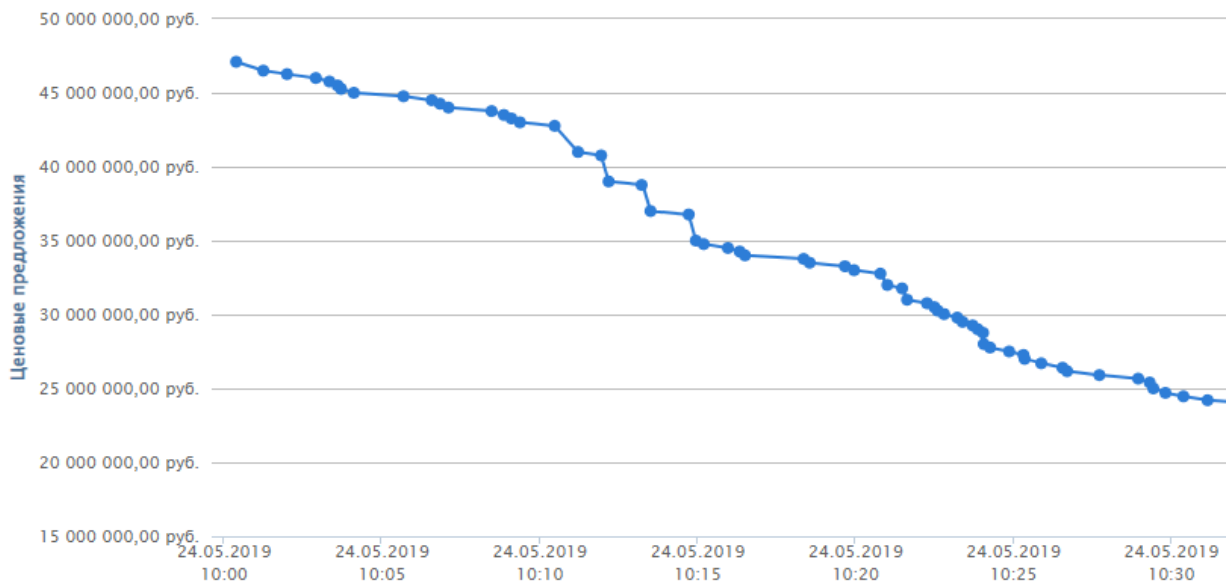

График 6. Ход торгов на аукционе с реестровым номером №31907838328. Источник: РТС-Тендер.

Цена на аукционе падала неестественно быстро. В материалах дела № 054/01/11-1272/2019 Новосибирского УФАС России указано следующее: "Фактически, по мнению антимонопольного органа ООО «СКС» и ООО «Коммунальные системы» при участии в аукционе электронной формы (извещение №31907838328) применялась так называемая схема «таран», то есть схема, при которой происходит максимальное снижение цены по антиконкурентному соглашению сторон. Однако в действиях Ответчиков отсутствуют признаки подобной схемы." Фирмы все же были признаны участниками картеля, хоть и в данном конкретном аукционе не применялась мошенническая схема, градиентный бустинг отметил этот аукцион как подозрительный на картель (график 7)

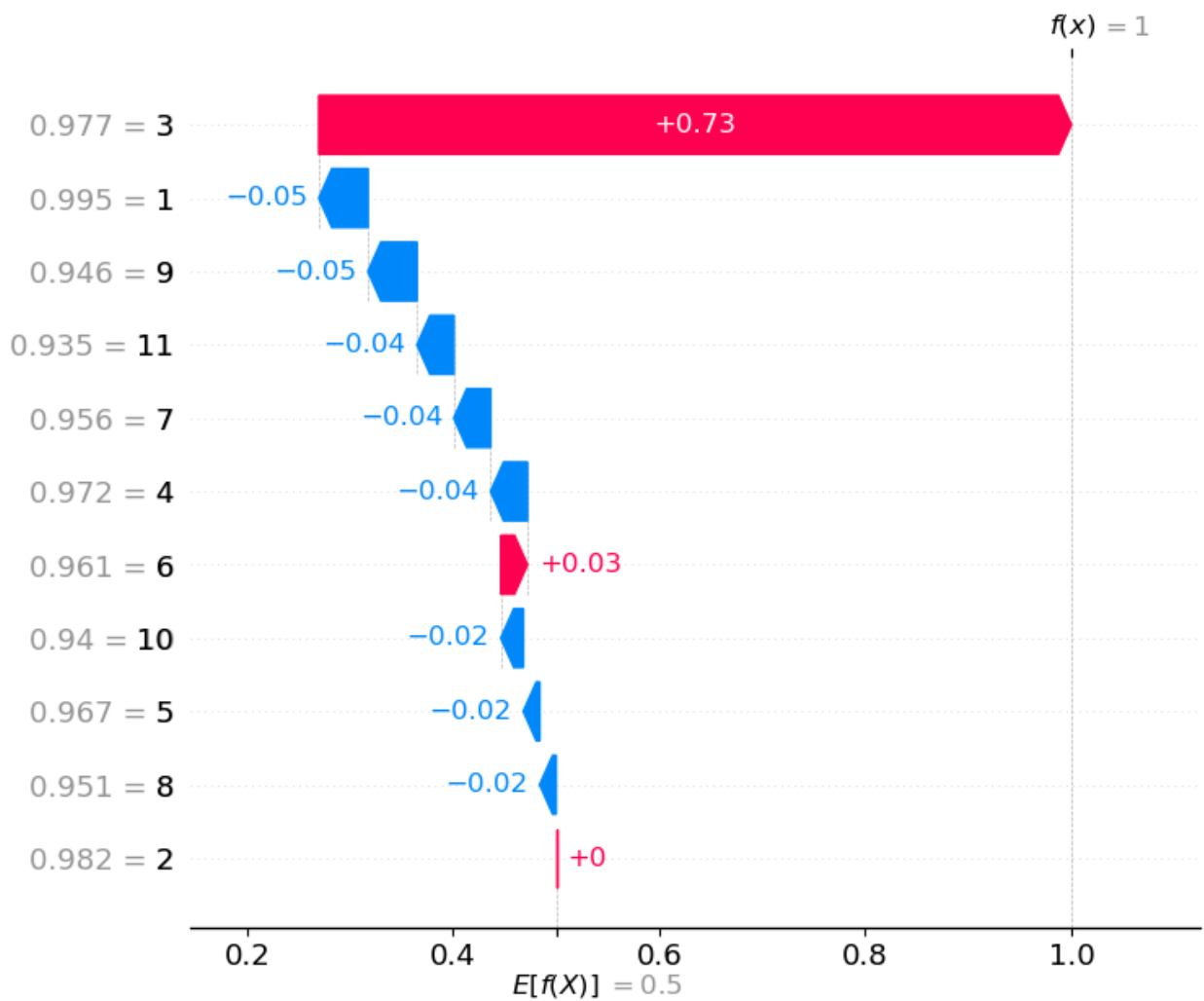

График 7. Интерпретация процесса решения градиентного бустинга. Ход торгов на аукционе с реестровым номером №31907838328. Источник: РТС- Тендер.

Ну и напоследок пример честного аукциона с точки зрения модели градиентного бустинга (график 8).

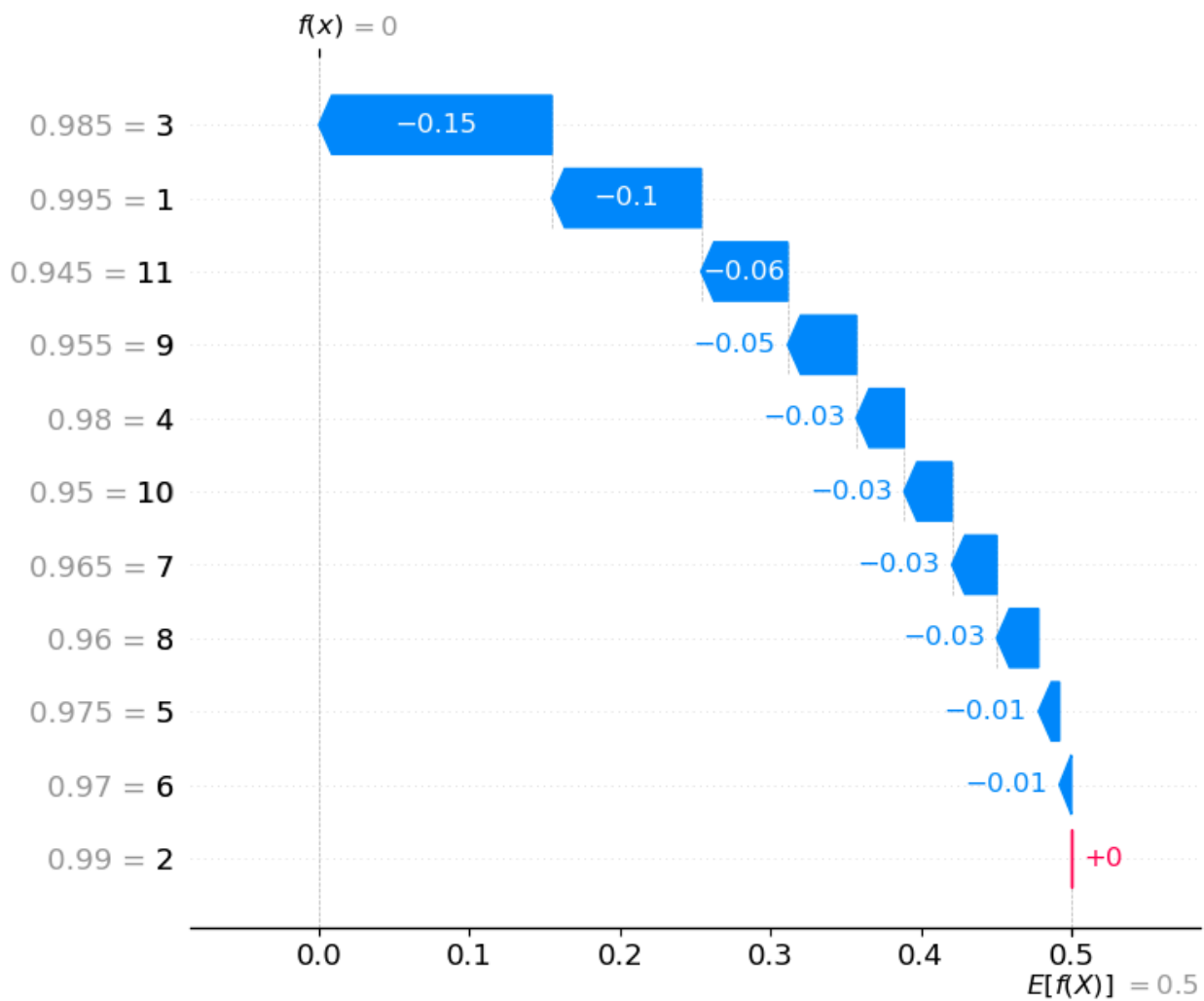

График 8. Интерпретация процесса решения градиентного бустинга для честного аукциона.
Источник: РТС- Тендер.

Видно, что каждая ставка отличалась от предыдущей ровно на 0.5%, то есть, все предлагали наименьшее снижение цены.

Можно построить усредненный график для всех аукционов сразу (график 9)

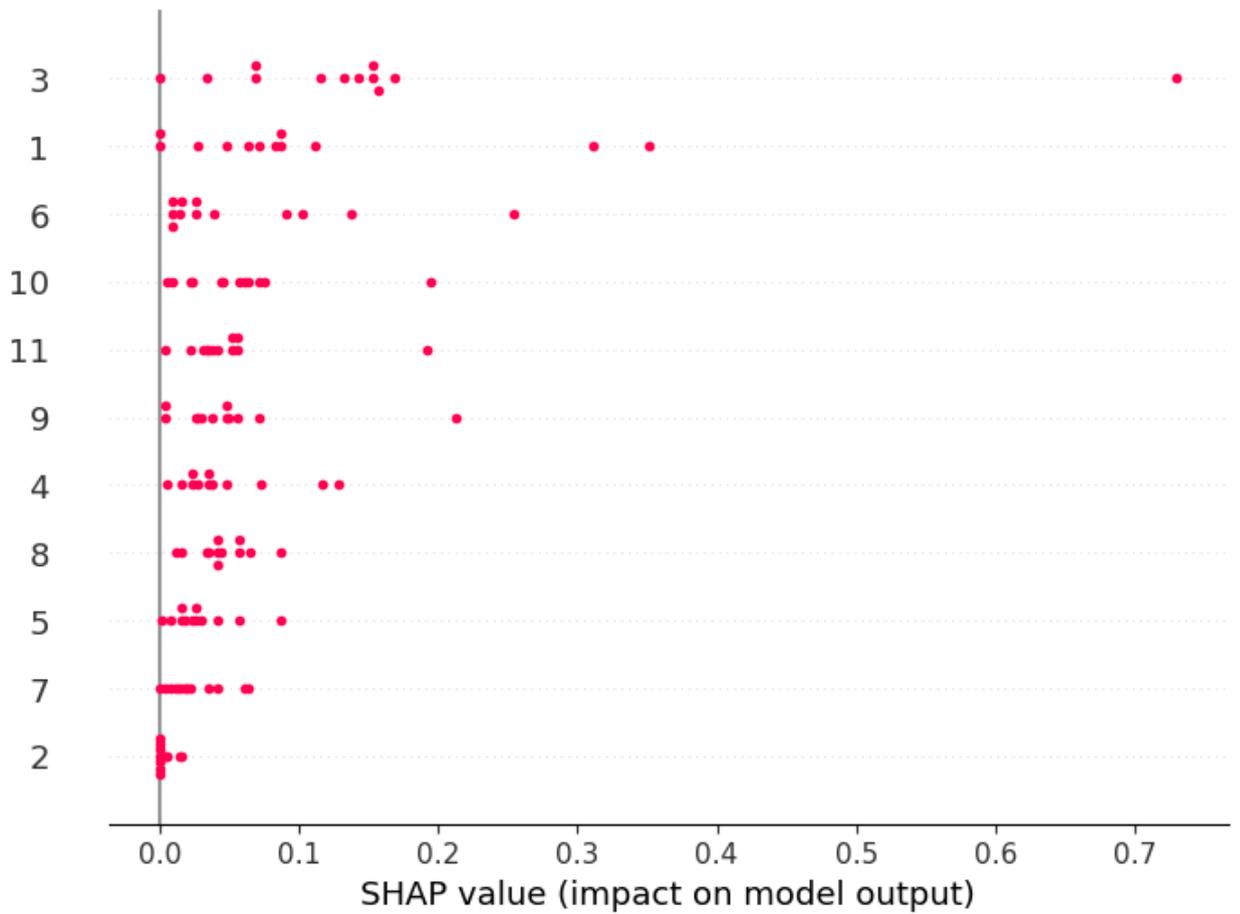

График 9. Усредненный вектор Шепли по всем аукционам. Источник: Расчеты автора.

По оси X представлено значение Шепли по модулю, по оси Y порядковый номер ставки. Каждая точка является влиянием конкретной ставки на конкретном аукционе на предсказание наличия картеля. Из графика видно, что первая, третья, шестая и четвертая ставки довольно часть помогали модели предсказать класс аукциона.

# ВЫВОДЫ

Алгоритмы машинного обучения обладают высоким потенциалом для поиска и раскрытия правонарушений, связанных с созданием картельных сговором. Точность градиентного бустинга превышает точность случайного угадывания, что говорит о подтверждении поставленной гипотезы. Качество работы модели может быть оценено по-разному, но, существует явный потенциал для обнаружения картельного сговора с помощью методов машинного обучения. Модель требует лишь несколько ставок для выдачи предсказания, что довольно сильно упрощает и ускоряет ее потенциальное развертывание. С помощью вектора Шепли удалось подтвердить результаты симуляционного процесса, а именно- смоделировать теоретическое представление о поведении конкурентной фирмы на открытом аукционе. Такая модель поможет сотрудникам антимонопольного органа ускорить и автоматизировать процесс мониторинга госзакупок и расширить его от уровня выборочных проверок до обязательной проверки каждого из проведенных тендеров на наличие подозрений о совершении незаконной деятельности. Основным недостатком подхода является невозможность предотвратить мимикрирование картельных фирм под честные. Как только информация о конкретном способе работы алгоритма станет известно картелю, ему ничего не помешает подавать всегда самое низкое предложение снижения цены. Тем не менее, если предположить будущую возможность модификации такой модели, например, добавление к величине ставки время ее подачи и превращения одномерного временного ряда в двухмерный, использование нейронных сетей вместо простой модели градиентного бустинга, то картельным фирмам было бы тяжело найти такое поведение, которое могло бы помочь сделать их похожими на конкретных игроков.

# СПИСОК ЛИТЕРАТУРЫ

Konstantin Efimov[3]

# Detecting collusion in procurement auctions[4]

The study aimed at detecting cartel collusion involved analyzing decisions of the Russian Federal Antimonopoly Service and data on auctions. As a result, a machine learning model was developed that predicts with 91% accuracy the signs of collusion between bidders based on their history after dividing 40 auctions into test and training samples in a 30/70 ratio. Decomposition of the model using the Shepley vector allowed the interpretation of the decision-making process. The behavior of 'honest' companies in auctions was also studied, confirmed by independent simulation validation.

Key words: cartels, electronic auctions, public procurement.

JEL: C57, L41, C45.

---

[3] Konstantin Dmitrievich Efimov - postgraduate student, National Research University Higher School of Economics (St. Petersburg, 190121, Russia)
[4] This work was supported by HSE University (Basic Research Program).